\documentstyle[aps,prb]{revtex}

\draft

\begin{document}
\input psfig

\title{Some Global Properties of the Attractive Hubbard Model\\ 
in the Superconducting 
Phase: T-Matrix Approximation in 2D}

\author{S. Schafroth}
\address{
Physik--Institut der Universit\"at Z\"urich,\\
Winterthurerstrasse 190, CH--8057 Z\"urich, Switzerland.
}
\author{J. J. Rodr\'{\i}guez-N\'u\~nez}
\address{Universidad Federal Fluminense, Instituto de F\'{\i}sica,\\
Av. Litor\^anea S/N, Boa Viagem, 24210-340 Niter\'oi, Brazil.}
\author{H. Beck}
\address{Universit\'e de Neuch\^atel, Institut de Physique,\\
Rue A.L. Breguet 1, CH--2000 Neuch\^atel, Switzerland.}

\date{\today}
\maketitle

\begin{abstract}\noindent
We have applied the Fast Fourier transform (FFT), which allows to 
compute efficiently convolution sums,   
to solve the set of self--consistent
T--matrix equations
to get the Green function of the 
two dimensional attractive--U Hubbard model
below $T_c$, extending previous calculations of the same authors. 
Using a constant order parameter $\Delta(T)$, we calculated $T_c$
as a function of electron density and interaction strength $U$.
These global results deviate from the BCS behavior remarkably.
\end{abstract}




\section*{}

Although the Hubbard model\cite{Hubbard} 
is the most simple model to 
describe correlated electron behavior 
in a solid, the mathematical treatment
is far from trivial. Many attempts have 
been made to understand the phase
diagram. A fully understood 
Hubbard model might form the basis of understanding
correlated electron systems as much as the Ising model
did for understanding critical 
phenomena in magnetism. The attractive--U
Hubbard model might play an 
important role in understanding high--temperature
superconductivity and has been 
attracting much attention in the past few years.
We have implemented the T--matrix approximation which 
goes beyond the usual mean--field approximation and
becomes exact in the dilute limit, 
i.e. where only two--particle interactions
take place\cite{5}.

We consider the attractive--U 
Hubbard model in two dimensions on a square lattice
(lattice constant $a$)\cite{MRR}

\begin{equation}
H = \sum_{{\bf k},\sigma}
    \xi_{\bf k} c_{{\bf k}\sigma}^\dagger  c_{{\bf k}\sigma} +
    U \sum_{{\bf k},{\bf k'},{\bf q}}
       c_{{\bf k}+{\bf q}\uparrow}^\dagger
       c_{-{\bf k}\downarrow}^\dagger
       c_{-{\bf k'}\downarrow}
       c_{{\bf k'}+{\bf q}\uparrow}
\end{equation}

\noindent with band energy 
$\epsilon_{{\bf k}} = -2 t (cos k_xa + cos k_ya)$
and on--site attraction $U < 0$, $\xi_{\bf k} = \epsilon_{\bf k} - \mu$,
$\mu$ is the chemical potential, and $t$, the 
hopping of electrons between nearest neighbour sites, determines
the energy unit. The creation (annihilation) operators for an 
electron with momentum ${\bf k}$ and spin $\sigma$ are denoted by
$c_{{\bf k}\sigma}^\dagger$ ($c_{{\bf k}\sigma}$).

The T--matrix (effective 
interaction) is the sum of 
particle--particle ladder diagrams with
the smallest number of closed fermion 
loops. In the low--density
limit, these are the 
dominating terms of the perturbation expansion in
terms of the interaction $U$.  
For the Hubbard model, where we have
only on--site interactions, the T--matrix approximation leads to a set
of self--consistent equations 
for the one-particle Green function \cite{Fresard} in 
the normal phase

\begin{equation}\label{Green}
G({\bf k},i\omega_n) = [i\omega_n - \xi_{\bf k} +
                        \Sigma({\bf k},i\omega_n)]^{-1},
\end{equation}

\noindent where the diagonal self--energy term

\begin{equation}\label{Sigma}
\Sigma({\bf k},i\omega_n) = {1\over {\beta N}} \sum_{{\bf q},m}
   T({\bf q},i\epsilon_m) G({\bf q}-{\bf k},i\epsilon_m-i\omega_n),
\end{equation}

\noindent depends on the T--matrix

\begin{equation}\label{T-Matrix}
T({\bf q},i\epsilon_n) = {-U \over {1+U\chi({\bf q},i\epsilon_n)}},
\end{equation}

\noindent which is a 
simple function of the independent pair--susceptibility

\begin{equation}\label{Chi}
\chi({\bf q},i\epsilon_n) = {1\over {\beta N}} \sum_{{\bf k},m}
   G({\bf k},i\omega_m) G({\bf q}-{\bf k},i\epsilon_n-i\omega_m).
\end{equation}

\noindent Here,
$\omega_n = (2n+1){\pi \over \beta}$ and
$\epsilon_n = 2n{\pi \over 
\beta}$ are the fermionic and bosonic Matsubara
frequencies, $N=N_xN_y$ where 
$N_x$ and $N_y$ are the grid dimensions
in ${\bf k}$--space and $\beta 
\equiv 1/T$ is the inverse temperature.
We fix the chemical potential 
by the electron density (one spin direction):

\begin{equation}\label{Density}
n(\beta,\mu) = {1 \over \beta N} \lim_{\eta \rightarrow 0^+}
 \sum_{{\bf k},n} G({\bf k},i\omega_n) e^{i\omega_n\eta}.
\end{equation}

To go below $T_c$, we introduce a 
constant order parameter 
$\Delta=|\Delta|$   
into the
Green function following the usual $2\times 2$ Nambu matrix formalism
and get for
the diagonal part the expression (${}^\ast$ means complex conjugate)
\begin{equation}\label{Green-with-Delta}
G({\bf k},i\omega_n) = {\Gamma^\ast({\bf k},i\omega_n)
 \over
|\Gamma({\bf k},i\omega_n)|^2 + |\Delta|^2}
\end{equation}

\noindent where 

\begin{equation}
\Gamma({\bf k},i\omega_n)=i\omega_n-\xi_{\bf k}+\Sigma({\bf k},i\omega_n).
\end{equation}

\noindent Equation (\ref{Green-with-Delta}) reduces to the usual
BCS Green function when the self--energy
term is set to zero (or to the Hartree shift). $\Delta$ 
is our
approximation for  $\Sigma_{12}({\bf k,i\omega_n})$, the off-diagonal 
self-energy. 
The order parameter $\Delta(T)$ is determined by

\begin{equation}\label{Gap-Equation}
{1 \over U} = {1 \over \beta N} \sum_{{\bf k},n}
{1 \over
|\Gamma({\bf k},i\omega_n)|^2+|\Delta|^2}
\end{equation}

\noindent which closes the set of equations.


To solve the set of equations (\ref{Sigma}), (\ref{T-Matrix}),
(\ref{Chi}), (\ref{Density}), (\ref{Green-with-Delta}) and
(\ref{Gap-Equation}) we apply the following
scheme:

\begin{enumerate}
\item Start by calculating $G_0({\bf k},i\omega_n)$, i.e. the Green
function for the free system ($\Sigma=0$).
A suitable initial value for $\mu$ and $\Delta$ must
be given ($\mu=-3.5$ and $\Delta=0.5$ are reasonable values for
$T=0.1$, $U=-4$ and $n=0.1$).

\item \label{Iteration-Start}
Calculate $\chi({\bf q},i\epsilon_m)$, $T({\bf q},i\epsilon_m)$ and
$\Sigma({\bf k},i\omega_n)$ using equations (\ref{Chi}), (\ref{T-Matrix})
and (\ref{Sigma}).

\item At this point, we need an improved estimation
of $\mu$ and $\Delta$. To get
a stable iteration scheme, both parameters must be adjusted simultaneously.
We do that by searching a solution $(\mu,\Delta)$ for equations (\ref{Density})
and (\ref{Gap-Equation}) using a Newton algorithm. 
For technical purposes we neglect the dependence on 
$\mu$ and $\Delta$ for the self--energy to 
be able to numerica

lly calculate the partial derivatives needed for 
the Newton algorithm.  
Because of that approximation we can not take the new
estimate of $(\mu,\Delta)$ directly.
Instead we go only a small step, in the $\mu-\Delta$
plane, from the current point to the new point (about one third of the total
distance).

\item  \label{Iteration-End}
Calculate an improved Green function using the new parameters $\mu$ and
$\Delta$.

\item Repeate steps \ref{Iteration-Start} to \ref{Iteration-End} until the
electron density has reached its desired value within a given tolerance.
\end{enumerate}

In order to obtain results which are independent of finite size, 
one should use at least some $10^3$
Matsubara frequencies and a grid of $30\times 30$ lattice points. The above
scheme works in principle but a closer 
look at the equations for $\chi$ and $\Sigma$ shows that the straight 
forward implementation of these equations does not work in practice. This is
due to the 4--fold loops which would occur in the computer program. Suppose
we use 2000 Matsubara frequencies and a $30\times 30$ grid. Then we have to
carry out for every grid point and every frequency the double sum over all
frequencies and all grid points. This are of the order 
$(30^2 \times 2000)^2 = 3.24\times 10^{12}$ complex operations. Even one of
the fastest super--computers would need one to several hours to make one
iteration step.

Since the frequency and momentum summations are convolutions,
we evaluate them using the fast Fourier transform.
The transforms ${\bf k} \rightarrow {\bf x}$ and
${\bf k} \leftarrow {\bf x}$ are the usual
ones and we do not elaborate them any further. The transforms from
$\tau \rightarrow i\omega$ and $\tau \leftarrow i\omega$ are described in more
detail. In the following, the notation

\begin{equation}\label{DefFFT}
 {\cal FFT}_M[F(x_j)]_n = {1\over \sqrt M} 
    \sum_{j=0}^{M-1} {e^{-2\pi i {jn\over M}}} F(x_j)
\end{equation}

\noindent and

\begin{equation}\label{DefFFTInverse}
 {\cal FFT}^{-1}_M[F(x_n)]_j = {1\over \sqrt M}
    \sum_{n=0}^{M-1} {e^{2\pi i {nj\over M}}} F(x_n)
\end{equation}

\noindent is used. The Matsubara frequencies are slightly redefined to be more
suitable (having non--zero indices) for numerical work and read:

\begin{equation}
  \omega_n = (2n+1-M){\pi\over\beta},\ \ \ \epsilon_n = (2n-M){\pi\over\beta}.
\end{equation}

\noindent We discretize the integral

\begin{equation}
  G(i\omega_n) = \int_0^\beta d\tau e^{i\omega_n\tau} G(\tau)
\end{equation}

\noindent by writing $\tau_j = j\Delta\tau,\  j=0\dots M-1$ where
$\Delta\tau = {\beta\over M}$ and $M$ is the number of Matsubara
frequencies used. We obtain

\begin{equation}\label{Gomega}
   G(i\omega_n) = {\beta\over\sqrt M} {\cal FFT}^{-1}_M[
   e^{i\pi j({1\over M} - 1)} G(\tau_j)]_n.
\end{equation}

\noindent The phase factor $e^{i\pi j({1\over M} - 1)}$ 
arises because of the
fermionic frequencies $\omega_n$ 
and the shift in the definition of
$\omega_n$. Similarly one can 
define the transforms for the bosonic
frequencies as

\begin{equation}
   X(i\epsilon_n) = {\beta\over\sqrt M} {\cal FFT}^{-1}_M[
   e^{-i\pi j} X(\tau_j)]_n,
\end{equation}

\noindent where $X$ is either $\chi$ or 
$\Sigma$. We can now rewrite equations
(\ref{Chi}) and (\ref{Sigma}) to read:

\begin{equation}\label{Chi-FFT}
   \chi(i\epsilon_n) = {\beta\over\sqrt M} {\cal FFT}^{-1}_M[
   e^{-i\pi j} G^2(\tau_j)]_n
\end{equation}
\noindent and
\begin{equation}\label{Sigma-FFT}
   \Sigma(i\omega_n) = -{\beta\over\sqrt M} {\cal FFT}^{-1}_M[
   e^{i\pi j({1\over M}-1)} T(\tau_j)G(-\tau_j)]_n.
\end{equation}

\noindent These expressions are also well suited for parallel
machines since the 3D--FFT (two space and one imaginary time
dimension) can be decomposed into parallel processes.

We also need to calculate the electron 
density, but evaluating (\ref{Density})
numericaly is not possible. To remove 
the limit $\eta \rightarrow 0^+$, we
make use of the definition of the Green function

\begin{equation}
  G(\tau) = \langle c^\dagger(\tau) c(0)\rangle,\ \ \ \tau > 0
\end{equation}

\noindent and 

\begin{equation}
  G(-\tau) = \langle c(0)c^\dagger(-\tau)\rangle =
             - 1 + \langle c^\dagger(-\tau) c(0)\rangle, \ \ \ \tau > 0.
\end{equation}

\noindent Taking the sum $G(\tau) + G(-\tau)$ and letting
$\tau \rightarrow 0^+$ we get

\begin{equation}
  n = {1 \over 2} [G(0^+) + G(0^-)] + {1 \over 2} = G(0) + {1 \over 2}.
\end{equation}

\noindent Therefore, the electron density now reads

\begin{equation}\label{density}
  n(\beta,\mu) = {1 \over 2} + {1\over \beta N}
                 \sum_{{\bf k},n} G({\bf k},i\omega_n),
\end{equation}

\noindent an expression which can be evaluated easily. A similar correction
for $G(\tau=0)$ must be applied in equations (\ref{Chi-FFT}) and
(\ref{Sigma-FFT}). Our Eq.(\ref{density}) generalizes Eq.(3.1.2) of 
Mahan's\cite{M}.

We have compared $n(\beta,\mu)$ for
the cases $\Sigma=0$ and $\Sigma=nU/2$ (the Hartree shift) with the
exact results and concluded that a grid of $32\times 32$ lattice sites
and 2048 frequencies is a lower limit for $U=-4$ and temperatures down
to $T=0.1$. For larger $|U|$ and/or lower temperatures one should
increase the number of frequencies.


So far we have evaluated the Green function at the Matsubara points.
Normally, we are really interested in the Fourier transform of the
retarded real--time Green function which is a function of real
frequency. In principle, we get this function by analytic continuation
of the complex frequency Green function from the Matsubara points to
the real axis. Since the resulting integral equations would be more
complicated (involving integrals over Fermi and Bose distribution functions) 
than the discrete frequency summations, we first calculate
the Green function at the Matsubara points as described in the
preceeding section. Then we continue to the real frequency axis by
fitting a rational function (M--point Pad\'e approximant) to the
calculated values \cite{PadeApprox}. The dynamical properties of the 
attractive Hubbard model in the superconducting phase is discussed in 
Ref.\cite{10}. Here, we concentrate on the global properties of the  
attractive Hubbard model.

The algorithm works as follows: Given a function $f(z_i)=u_i$ with values
$u_i$ at $M$ complex points $z_i,\ i=1,2,\dots,M$, the Pad\'e approximant
is defined as a ratio of two polynomials which can be written as a
continued fraction

\begin{equation}\label{Pade-Approximant}
 C_M(z) = {a_1 \over 1 + {\displaystyle a_2(z-z_1) \over \displaystyle 
           1 + \dots {
           \displaystyle a_M(z-z_{M-1}) \over \displaystyle 1}}}
\end{equation}

\noindent where the coefficients $a_i$ are to be determined so that

\begin{equation}
  C_M(z_i) = u_i, \ \ \ i=1,2,\dots,M
\end{equation}

\noindent which is fulfilled when the $a_i$ are given by the recursion

\begin{equation}
  a_i = g_i(z_i), \ \ \ \ g_1(z_i)=u_i, \ \ \ \ i=1,2,\dots,M
\end{equation}

\noindent and

\begin{equation}
  g_p(z) = {g_{p-1}(z_{p-1})-g_{p-1}(z) \over (z-z_{p-1})g_{p-1}(z)},
 \ \ \ \ p \ge 2.
\end{equation}

\noindent Once the coefficients $a_i$ are determined for a particular
function, the function values at a real frequency $\omega$ can be obtained
by setting $z=\omega+i\delta$ in (\ref{Pade-Approximant}), where $\delta$ is
used to remove unphysical peak structures due to finite size effects. The
choice $\delta=0.1\dots0.2$ is appropriate for our calculations. 
In reality, the choice of  $\delta$ is given by the precision with 
which the self-consistency in the Matsubara frequencies is 
done. This point has been discussed by Georges et al\cite{GKKR}. 
In this paper we limit the discussion of the results to the regime $T<T_c$,
since results for $T>T_c$ are reported elsewhere \cite{Results1}. 

\begin{figure}[h]
\centerline{\psfig{figure=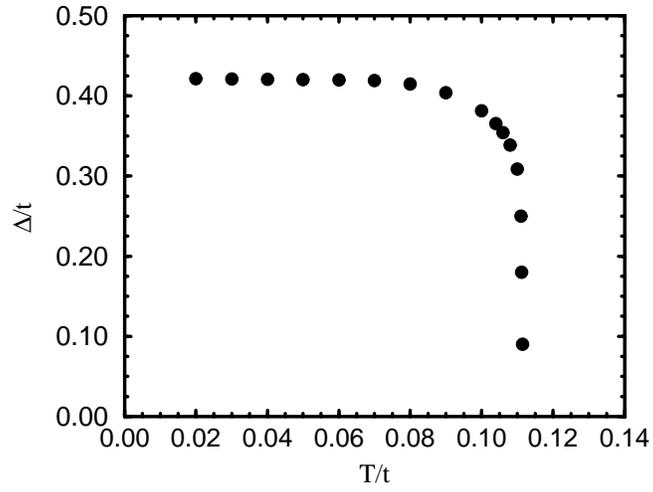,height=70mm,rheight=67mm}}
\caption{\label{Delta-T}
Order parameter $\Delta$ versus temperature for density $n=0.1$ and interaction
strength $U=-4$.}
\end{figure}

Figure \ref{Delta-T} shows the temperature dependence of the order
parameter for $n=0.1$ and $U=-4t$. The system size is 32 by 32 grid
points in ${\bf k}$--space and 1024 Matsubara frequencies. The order
parameter shows a very sharp drop to zero (compared with BCS behavior)
which indicates the strong influence of fluctuations for all temperatures.
$\Delta(T=0)$ is nearly twice as large as the corresponding
$\Delta_{BCS}$ whereas the critical temperature is much lower then
$T_c^{BCS}$ by more than a factor three. For example, $\Delta(0)/T_c 
\approx 3.86$, which is more than twice the $BCS$ ratio ($\approx 
1.76$). This implies that we are not in the weak coupling limit, opposite 
to the case of Mart\'{\i}n-Rodero and Flores\cite{12} who find the 
BCS universal ratio in second order of perturbation for the 
continuous Hubbard model. The value of $\Delta(T) = 0$ defines 
the critical temperature, $T_c$. In order to calculate $T_c$ we 
have drawn a straight line close to the sharp drop of 
the order parameter. Near the critical temperature we had 
some convergence problems. However, the evaluation of 
$T_c$ is precise because it is calculated as described 
previously. 

\begin{figure}[h]
\centerline{\psfig{figure=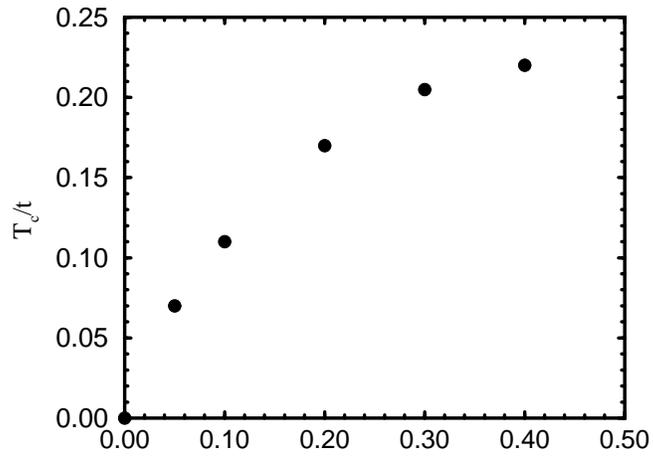,height=70mm,rheight=67mm}}
\caption{\label{Tc-n}
Critical temperature $T_c$ versus electron density $n$ ($n=0.5$ corresponds
to half--filling) for an interaction strength $U=-4$.}
\end{figure}

The density dependence of $T_c$ is plotted in Figure \ref{Tc-n}. It is
well known, that in the strong coupling limit the attractive--U Hubbard
model close to half--filling can be mapped to a Heisenberg model which shows
no Kosterlitz--Thouless transition \cite{Heisenberg}. It is therefore
expected, that the critical temperature near half-filling is reduced
even in the weak-- and intermediate--coupling regime.  The lack of
this property in our results can be attributed to the
neglect of the particle--hole channel (charge fluctuations) in the T--matrix. 
Then, we should restrict ourselves to low densities where the T--matrix 
approach is indeed valid. To treat higher densities, we should implement,  
for example, the $FLEXC$ approach\cite{6} goal which we are pursuing at this 
moment for the attractive Hubbard model. Here, we mention that the 
T--matrix and $FLEXC$ approaches are conserving in the Kadanoff-Baym 
sense\cite{5}.  
Denteneer et al\cite{Denteneer} obtained the critical temperature by
calculating the helicity modulus associated with a wavelike distortion
($\Delta_j = |\Delta| exp(2i{\bf qr}_j)$) of the order parameter in
the BCS approximation as a function of temperature and subsequent
comparison with the Kosterlitz--Thouless relation between critical
temperature and helicity modulus. But they miss the logarithmic
drop to zero when approaching half--filling.

\begin{figure}[h]
\centerline{\psfig{figure=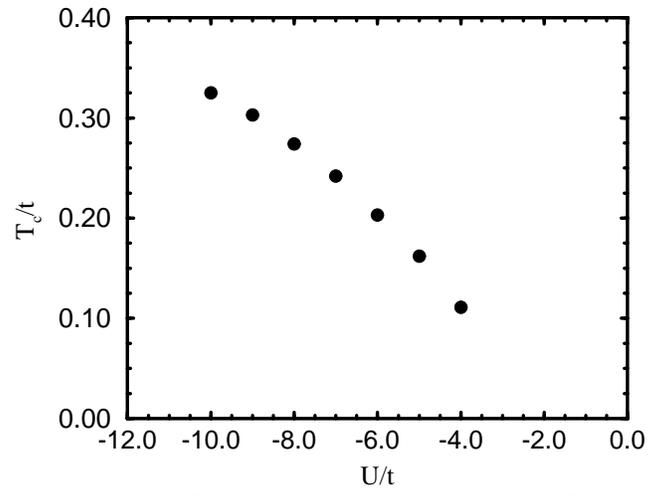,height=70mm,rheight=67mm}}
\caption{\label{Tc-U}
Critical temperature $T_c$ for various interaction strengths $U$ for constant
density $n=0.1$.}
\end{figure}

Figure \ref{Tc-U} shows the critical temperature as a function of interaction
strength. In contrast to the BCS behavior (which shows an exponential increase
of $T_c$ small $|U|$ and goes linear in 
$|U|$ for larger values), the increase of $T_c$ is reduced drastically. 
The expected decrease of $T_c$ for large $|U|$ can not be observed for the
values of $U$ treated here. The
expectation is based on the fact, that for strong attraction and near to
half--filling the model can be
mapped onto a pseudo--spin model with effective interaction constant
$J=4t^2/|U|$, resulting in a $T_c$ decreasing with increasing $|U|$. 
The behavior we observe for $T_c/t$ vs $U/t$ is similar to the one 
obtained by Nozi\`eres and Schmitt-Rink\cite{NSR} using the Thouless' 
criteria in the normal phase. Most likely we have to include 
additional fluctuations in the 
off-diagonal self-energy\cite{SRN}. Contrary to the 
normal state calculations where, for $\rho = 0.1$, we could not 
obtain convergence for $|U|~>~4.0$, now in the superconducting 
phase the program is stable for large values of $|U|$.

In conclusion, we have calculated the critical temperature of the
negative--U Hubbard model within the T--matrix approximation. The
expected smooth cross--over from BCS to Bose--condensation can not be
fully observed in the parameter space studied in this paper.
Although the exponential-to-linear increase of $T_c^{BCS}$ with
growing $|U|$ is reduced drastically, an optimal critical temperature
has not been found. Perhaps we should fully 
go beyond $BCS$ in the off-diagonal 
self-energy, as it has been 
done by Shafroth and Rodr\'{\i}guez-N\'u\~nez\cite{SRN}. 
In this work\cite{SRN}, the 
authors have studied the dynamical properties of the attractive 
Hubbard model in presence of {\it double fluctuations}. 
As previously discussed, we have obtained that 
$\Delta / T_c$ is more than twice the BCS value.
To get $T_c = 0$ at half--filling we must include charge fluctuations.
Investigations along these lines is under way\cite{RNS}. The value 
$\Delta(0)/T_c \approx 3.86$ and the temperature behavior of 
$\Delta(T)$ at $U/t = - 4.0$, $n = 0.1$ are very different with 
respect to $BCS$ results. These global results suggest 
that for $U/t = - 4.0$, $n = 0.1$ 
we are already in the intermediate coupling regine where 
correlations are important.

This work has been supported by the Swiss National Science Foundation 
and CNPq (Brazilian Agency). We would  
like to thank Prof. T. Schneider,  
Dr. M.H. Pedersen, Prof. R. Micnas and Dr. 
J.M. Singer for helpful and stimulating 
discussions. We thank Prof. Mar\'{\i}a D. Garc\'{\i}a 
Gonz\'alez for reading the 
manuscript.

\newpage
\end{document}